\documentclass[conference]{IEEEtran}
\IEEEoverridecommandlockouts

\usepackage{cite}
\usepackage{amsmath,amssymb,amsfonts}
\usepackage{algorithmic}
\usepackage{bm}
\usepackage{array}
\usepackage{graphicx}
\usepackage{textcomp}
\usepackage{xcolor}
\usepackage{multirow}
\def\BibTeX{{\rm B\kern-.05em{\sc i\kern-.025em b}\kern-.08em
    T\kern-.1667em\lower.7ex\hbox{E}\kern-.125emX}}
\begin{document}

\title{Adaptive Momentum Enhanced Distributed Multichannel Active Noise Control for Faster Convergence under Communication Delays
}

\author{\IEEEauthorblockN{Junwei Ji, Woon-Seng Gan, Boxiang Wang, Ziyi Yang, Haowen Li}
\IEEEauthorblockA{\textit{School of Electrical and Electronic Engineering, Nanyang Technological University, Singapore} \\
Email: JUNWEI002@e.ntu.edu.sg, ewsgan@ntu.edu.sg}
}

\maketitle

\begin{abstract}
Distributed multichannel active noise control (DMCANC) reduces the computational burden of centralized ANC systems by distributing processing tasks across multiple nodes, while requiring information exchange to achieve satisfactory global noise reduction. To improve robustness under communication delays, the auto-shrink step size mixed-gradients filtered reference LMS (ASSS-MGDFxLMS) algorithm has been proposed. However, the reduced step size inevitably slows convergence. In this work, an adaptive momentum term is introduced to accelerate convergence, where cosine similarity is used to evaluate the alignment between the instantaneous gradient and the momentum component and dynamically adjust the momentum parameter. This design accelerates convergence when the directions are consistent while preserving stability under delayed communication. Simulation results demonstrate that the proposed adaptive momentum ASSS-MGDFxLMS (AMAS-MGDFxLMS) algorithm achieves faster convergence than ASSS-MGDFxLMS while maintaining stable and effective noise reduction performance.

\end{abstract}


\section{Introduction}
By employing multiple secondary sources and error sensors, multichannel active noise control (MCANC) systems enable global noise reduction over large spatial regions \cite{Elliott2001SPAC}. To achieve satisfactory performance and effectively track dynamic noise characteristics and acoustic environments, adaptive algorithms are widely employed. Among them, the filtered reference least mean square (FxLMS) algorithm \cite{Morgan1980FXLMS} is one of the most classical approaches and has been extensively extended to multichannel configurations \cite{Kuo1999ANC}. In recent years, a wide range of FxLMS variants have been developed to address practical challenges encountered in real-world implementations \cite{lam2021ten}, such as output saturation \cite{guo2024survey}, spatial reduction \cite{su2025co,zhang2025spherical}, and virtual sensing \cite{tsuji2025novel,wang2025transferable}. Recently, advances in artificial intelligence (AI) have further stimulated the development of deep-learning-based ANC algorithms \cite{zhang2021deep,xie2024cognitive,pike2025dynamic}, offering promising potential for enhanced robustness and adaptability in complex and highly dynamic acoustic environments \cite{luo2025frequency}.

Conventional MCANC algorithms, such as the multiple error FxLMS (MEFxLMS) approach \cite{Elliott1987MEANC}, typically adopt a centralized architecture:in which a single processor performs multichannel adaptive processing using all reference and error signals. Although effective, the computational load grows rapidly with the number of channels, leading to high hardware cost and limited real-time scalability \cite{Douglas1999FXLMS,shi2023computation}. To reduce complexity, various efficient algorithms have been proposed, including partial-update schemes \cite{shi2018partial}, mixed-error methods \cite{murao2017mixed}, and frequency-point selection techniques \cite{lian2025computational}. As an alternative, decentralized strategies distribute computations across multiple local controllers, each updating its control filter using only local error signal \cite{Pradhan2023DeMCANC,Zhang2019Decentralized}. However, decentralized methods can degrade global attenuation and may even compromise stability due to inter-channel interference \cite{Elliott1994InteractionMANC}.

\begin{figure}[!t]
    \centering
    \includegraphics[width = 0.7\columnwidth]{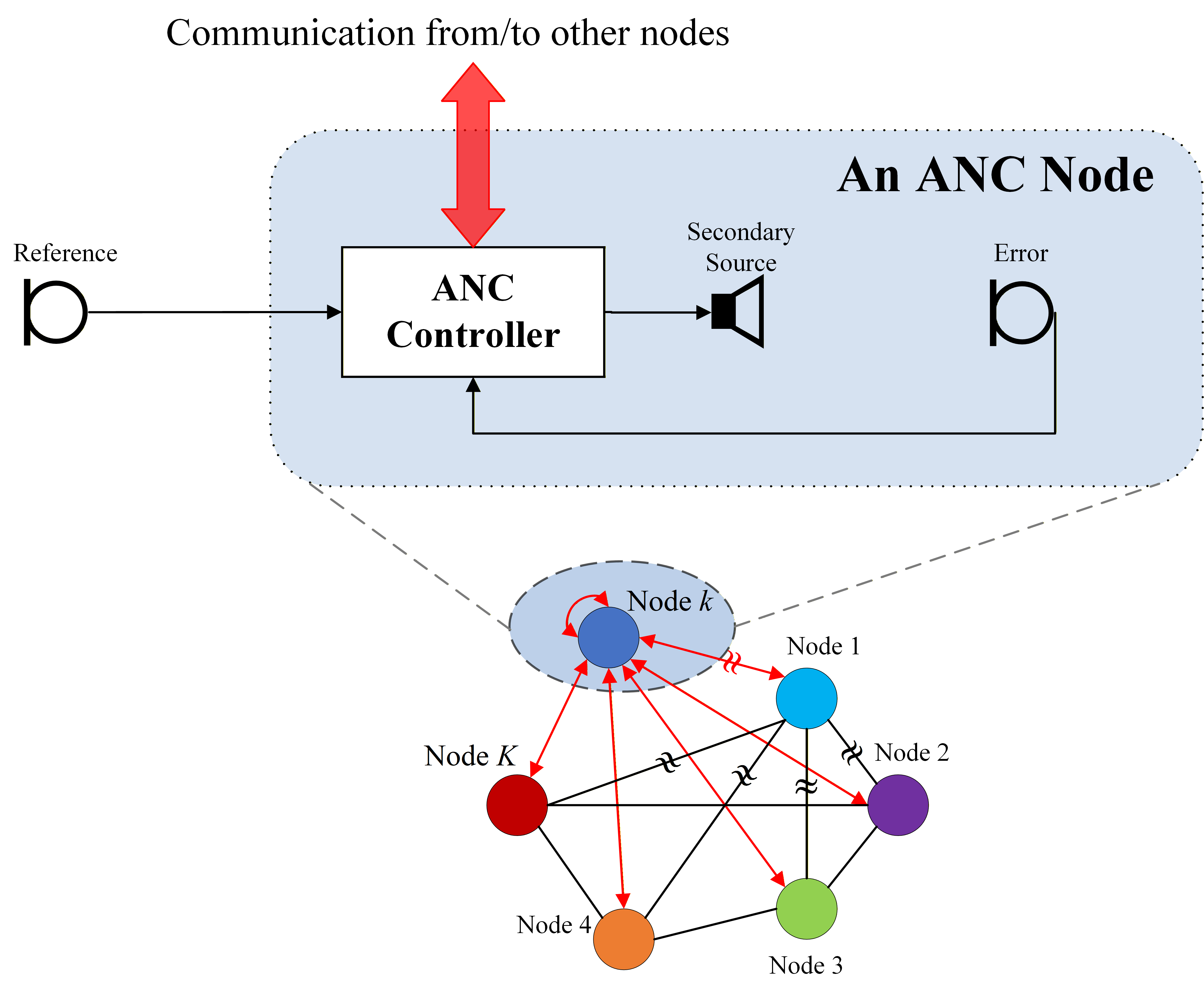}
    \caption{A general DMCANC network, where each ANC node consists of a secondary source, an error sensor, and an ANC controller.}
    \label{fig:DMCANCnets}
\end{figure}

Distributed MCANC (DMCANC) methods provide a promising alternative \cite{ferrer2015active}, where each node processes local signals while exchanging essential information to achieve global noise reduction \cite{Antonanzas2015Diffusion}. A typical DMCANC network is shown in \ref{fig:DMCANCnets}. The diffusion FxLMS (DFxLMS) algorithm was developed by employing topology-based combination rules \cite{Chu2019DiffusionANC,Chu2020DiffusionANC}, but its spatial-smoothing nature limits performance under asymmetric acoustic paths \cite{Chu2021Combination}. To address this issue, the augmented DFxLMS (ADFxLMS) algorithm and a bidirectional communication strategy were proposed to improve robustness and reduce communication overhead \cite{Li2023Distributed,Li2023AugmentedDiffusion}. Recent experimental studies have further validated DMCANC using in-chip communication platforms \cite{li2024experimental}.

However, these approaches are generally limited to ideal communication networks. Under communication latency, theoretical analysis has shown that DMCANC systems may become unstable \cite{Ji2025MGDFXLMS}. To address this issue, the auto-shrink step size mixed-gradients distributed FxLMS (ASSS-MGDFxLMS) algorithm was proposed, in which the step size at each node is automatically reduced according to the degree of communication delay \cite{Ji2025MGDFXLMS}. While this strategy enhances stability, the reduced step size inevitably slows down the convergence speed. To address this limitation, a momentum term is introduced to accelerate convergence. To avoid instability caused by a fixed momentum factor, cosine similarity is used to evaluate the directional consistency between the mixed gradient and the momentum term \cite{rashidi2020adaptive}, enabling adaptive adjustment of the momentum strength. Consequently, the proposed adaptive momentum ASSS-MGDFxLMS (AMAS-MGDFxLMS) improves DMCANC performance under communication latency by rapidly achieving satisfactory noise reduction while preserving system stability, demonstrating strong practical significance.

The remainder of this paper is organized as follows: Section~\ref{sec:review_asss} reviews the ASSS-MGDFxLMS algorithm. Section~\ref{sec:AMASMGDFxLMS} presents the proposed adaptive momentum strategy to improve the convergence speed of ASSS-MGDFxLMS under communication latency. Section~\ref{sec:sims} reports numerical simulations that demonstrate the effectiveness of the proposed method. Finally, Section~\ref{sec:concl} concludes the paper.

\section{The ASSS-MGDFxLMS Algorithm}
\label{sec:review_asss}
In practical deployments, DMCANC systems suffer from two critical challenges: inter-node acoustic crosstalk and non-ideal communication conditions, especially communication delays. The ASSS-MGDFxLMS algorithm was proposed to jointly address these issues and forms the basis of the present work.

\subsection{Mixed-gradients distributed FxLMS}
\label{subsec:mgdfxlms}
Consider a DMCANC system consisting of $K$ ANC nodes that share a common reference microphone. The control signal generated by the $k$th node is given by
\begin{equation}\label{eq:controlsignal}
    y_k(n) = \mathbf{w}_k^\mathrm{T}(n)\mathbf{x}(n), \quad k = 1,2,...,K,
\end{equation}
where $\mathbf{w}_k(n)=[w_{k,0}(n) \, w_{k,1}(n) \, \cdots \, w_{k,L_w-1}(n)]^\mathrm{T}$ and $\mathbf{x}(n)=[x(n) \, x(n-1) \, \cdots \, x(n-L_w+1)]^\mathrm{T}$ denote the $k$th control filter and reference signal vectors with the length of $L_w$, and $n$ is the discrete-time sample index. The control signal propagates through the secondary path to suppress the disturbance $d_k(n)$ at the error microphone. In addition to the anti-noise generated by the $k$th node, the error microphone also captures the anti-noise contributions from other nodes due to acoustic crosstalk. Consequently, the error signal at the $k$th node can be expressed as
\begin{equation}\label{eq:error}
    e_k(n) = d_k(n) - y_k(n) * s_{kk}(n)-\sum_{{m=1,m\neq k}}^{K}y_m(n)*s_{km}(n),
\end{equation}
where $*$ denotes linear convolution, $s_{kk}(n)$ represents the self-secondary path of the $k$th node, and $s_{km}(n)$ denotes the cross-secondary path from the $m$th secondary source to the $k$th error microphone.

Based on the gradient descent method, the instantaneous gradient of the local cost function at node $k$ with respect to the control filter $\mathbf{w}_k(n)$ is given by
\begin{equation}\label{eq:localgradient}
    \boldsymbol{\nabla}_k(n) = -2  [\mathbf{x}(n)*\hat{s}_{kk}(n)]\cdot e_k(n).
\end{equation}
where $\hat{s}_{kk}(n)$ denotes the estimated self-secondary path. Similarly, the local gradients of other nodes can be computed following \eqref{eq:localgradient}. Instead of transmitting the control filters as the conventional DMCANC algorithms, the local gradient of each node is shared. After receiving the gradients from other nodes, each node updates its global control filter as
\begin{equation}\label{eq:GCF}
    \begin{split}
            \mathbf{w}_k(n+1) = \mathbf{w}_k(n) - \mu\big[\boldsymbol{\nabla}_k(n) + \sum_{{m=1, m\neq k}}^K \boldsymbol{\nabla}_m(n)*c_{mk}(n)\big],   
    \end{split}
\end{equation}
where $\mu$ is the step size, and $c_{mk}(n)$ denotes the compensation filter introduced to approximate the mismatch between the self-secondary path and the cross-secondary path \cite{Ji2023Distributed}, thereby mitigating the crosstalk effect. Specifically, the relationship between the estimated secondary path and the compensation filter is given by
\begin{equation}\label{eq:compensate}
 \hat{s}_{mk}(n) = \hat{s}_{mm}(n) * c_{mk}(n), \; (m\neq k).
\end{equation}
These formulations constitute the mixed-gradients distributed FxLMS (MGDFxLMS) algorithm \cite{Ji2025MGDFXLMS}. By exchanging local gradient information rather than control filter coefficients, MGDFxLMS extends the conventional centralized MEFxLMS algorithm to a distributed framework, enabling effective suppression of inter-node crosstalk while maintaining performance comparable to centralized MCANC.

\begin{figure}[!t]
    \centering
    \includegraphics[width = 0.85\columnwidth]{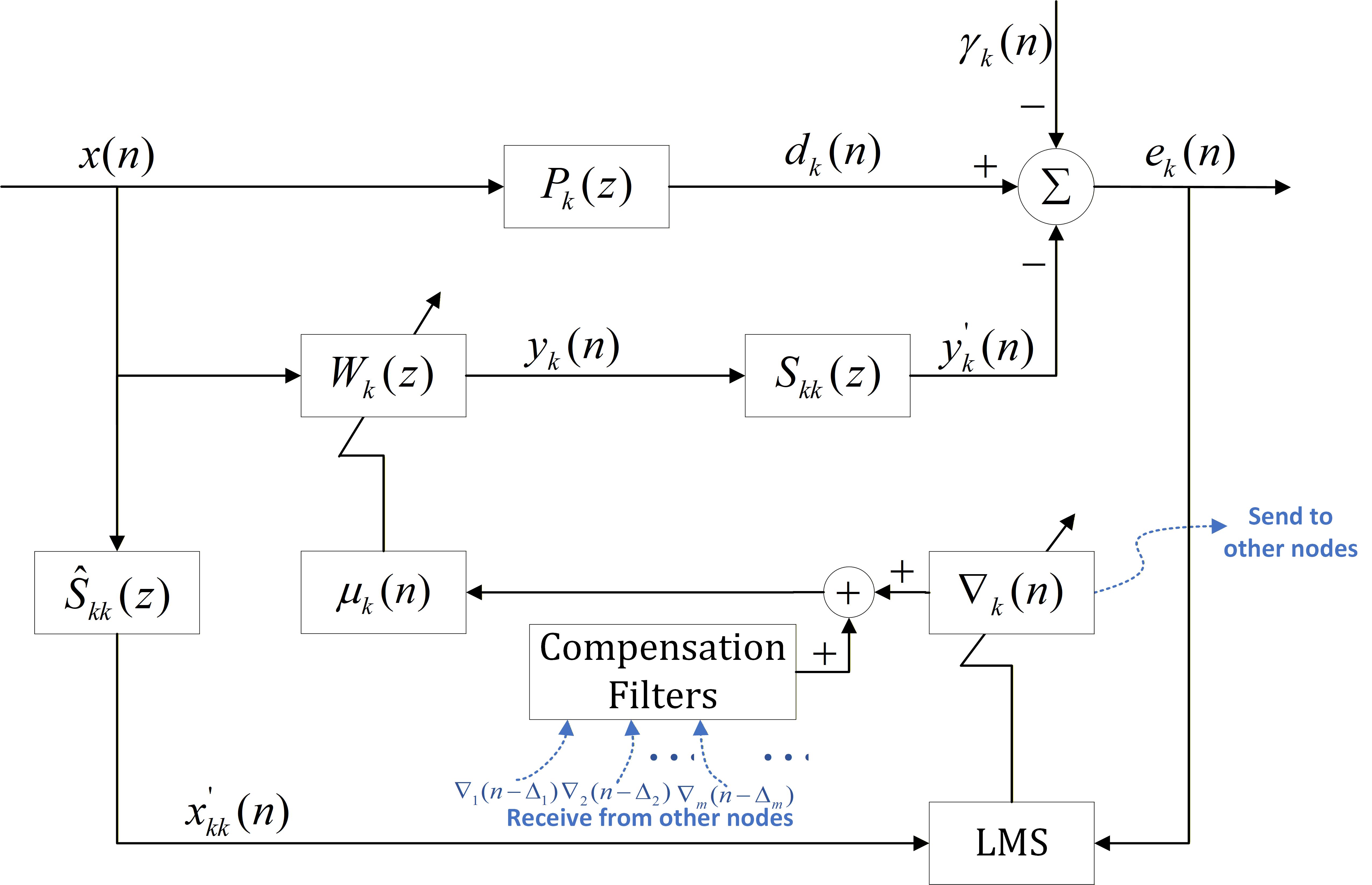}
    \caption{Block diagram of the ASSS-MGDFxLMS algorithm for the $k$th node, where $\gamma_k(n)$ denotes the interference generated by other nodes.}
    \label{fig:asss_mgdfxlms}
\end{figure}

\subsection{Auto-shrink step size (ASSS)}
\label{subsec:asssmgdfxlms}
In practical distributed ANC systems, communication delays are unavoidable and can significantly degrade the stability of the algorithm. To account for this effect, the update rule is modified by incorporating the delayed gradient from the $m$th node, denoted by $\Delta_m$, leading to
\begin{equation}\label{delayGCF}
    \begin{split}
           &\mathbf{w}_k(n+1) = \mathbf{w}_k(n)\\& - \mu\big[\boldsymbol{\nabla}_k(n) + \sum_{m=1,m\neq k}^K\boldsymbol{\nabla}_m(n-\Delta_m)*c_{mk}(n)\big]. 
    \end{split} 
\end{equation}
To mitigate the latency issue, an auto-shrink step size strategy is introduced, defined as
\begin{equation}\label{eq:asss_stepsize}
\mu_k(n) = \mu_0\exp(-{2\Delta}/{f}),
\end{equation}
where $\mu_0$ denotes the initial step size, $f$ is the sampling frequency, and
\begin{equation}\label{eq:maxdelay}
    \Delta = \max[\{\Delta_m \mid m=1,2,\cdots,K \;\text{and} \; m\neq k \}],
\end{equation}
represents the maximum communication delay among the received nodes. Consequently, the auto-shrink step size MGDFxLMS (ASSS-MGDFxLMS) algorithm is expressed as
\begin{equation}\label{eq:Genearlupdate}
    \begin{split}
           &\mathbf{w}_k(n+1) = \mathbf{w}_k(n)\\&- \mu_k(n)\big[\boldsymbol{\nabla}_k(n) + \sum_{m=1,m\neq k}^K\boldsymbol{\nabla}_m(n-\Delta_m)*c_{mk}(n)\big]. 
    \end{split}
\end{equation}
The block diagram of ASSS-MGDFxLMS is illustrated in Fig.~\ref{fig:asss_mgdfxlms}, where the local gradient of each node is transmitted and combined with the compensation filter to compute the global control filters and the auto-shrink step size strategy allows the step size value to be adaptive to communication delays to enhance the system's robustness. 

\section{Adaptive momentum method}
\label{sec:AMASMGDFxLMS}
Although the ASSS-MGDFxLMS algorithm significantly improves system stability under communication delays, this robustness is achieved at the expense of a reduced convergence speed. To alleviate this drawback, a momentum term is introduced to accelerate convergence.

The mixed gradient at the $k$th node is defined as
\begin{equation}\label{eq:mgd}
    \mathbf{g}_k(n) = - \mu_k(n)\big[\boldsymbol{\nabla}_k(n) + \sum_{m=1,m\neq k}^K\boldsymbol{\nabla}_m(n-\Delta_m)*c_{mk}(n)\big],
\end{equation}
and the corresponding momentum term is given by
\begin{equation}\label{eq:momterm}
           \mathbf{v}_k(n) = \mathbf{w}_k(n)-\mathbf{w}_k(n-1). 
\end{equation}
By incorporating the momentum term, the general update equation in \eqref{eq:Genearlupdate} can be modified as
\begin{equation}\label{eq:fmas_mgdfxlms}
           \mathbf{w}_k(n+1) = \mathbf{w}_k(n) +\mathbf{g}_k(n) + \beta\mathbf{v}_k(n),
\end{equation}
where $\beta$ denotes the momentum factor controlling the contribution of the momentum term. However, under communication latency, employing a fixed momentum factor may lead to instability, particularly in fluctuating networks. This motivates the adoption of an adaptive momentum strategy.

To dynamically regulate the momentum contribution, the cosine similarity \cite{rashidi2020adaptive} is employed to evaluate the directional consistency between the mixed gradient and the momentum term, defined as
\begin{equation}\label{eq:cossim}
    \rho_k(n) = \frac{\mathbf{g}^\mathrm{T}_k(n)\mathbf{v}_k(n)}{||\mathbf{g}_k(n)||\;||\mathbf{v}_k(n)||}.
\end{equation}
Since both $\mathbf{g}_k(n)$ and $\mathbf{v}_k(n)$ are high-dimensional vectors, the resulting cosine similarity values are typically small. To mitigate this issue, a power-law scaling is applied \cite{caograms}, leading to the following adaptive momentum factor:
\begin{equation}\label{eq:am}
    \beta_k(n) = \text{min}(\beta_0|\rho_k(n)|^p,\beta_0),
\end{equation}
where $0<p<1$ and $\beta_0$ is the maximum momentum factor. The parameter $\beta_0$ can be estimated under ideal network conditions \cite{shen2025multi}, while $p$ is selected according to the order of magnitude of $\rho_k(n)$. When the gradient and momentum directions are well aligned, a larger momentum factor is applied to accelerate convergence; otherwise, the momentum contribution is attenuated to enhance stability. Therefore, the update rule for the proposed adaptive momentum ASSS-MGDFxLMS (AMAS-MGDFxLMS) is expressed as
\begin{equation}\label{eq:amas_mgdfxlms}
           \mathbf{w}_k(n+1) = \mathbf{w}_k(n) +\mathbf{g}_k(n) + \beta_k(n)\mathbf{v}_k(n). 
\end{equation}
The pseudo-code of the AMAS-MGDFxLMS algorithm is summarized in Table~\ref{tab:algorithm}. By adaptively adjusting the momentum factor based on the directional consistency between the mixed gradient and the momentum term, the proposed method effectively accelerates convergence while preserving stability. The additional computational cost is approximately $4L_w+2$ multiplications and $6L_w-3$ additions per node at each iteration, excluding a few scalar nonlinear operations. Consequently, the AMAS-MGDFxLMS algorithm achieves rapid and reliable noise reduction performance under fluctuating network conditions.

\begin{table}[!t]
    \centering
    \caption{Pseudo-code of the AMAS-MGDFxLMS algorithm}\label{tab:algorithm}
    \begin{tabular}{l}
    \hline
    \textbf{Algorithm:} the AMAS-MGDFxLMS algorithm for the $k$th ANC nodes.\\
    \hline
    \textbf{Initialization:} Obtain estimated self-secondary path, $\hat{s}_{kk}(n)$, \\and compensation filters, $\mathbf{c}_{mk}(n)$.\\
    \textbf{Input:} The reference signal $x(n)$; The error signal $e_k(n)$; \\Received gradients from other nodes $\boldsymbol{\nabla}_m(n-\Delta_m)$, ($m\neq k$).\\
    \textbf{Output:} The control signal $y_k(n)$; The $k$th local gradients $\boldsymbol{\nabla}_k(n)$.\\
    \textbf{While} True \textbf{do}\\
     /*Combine gradients to obtain global control filter */\\
     ~~~$\Delta\leftarrow\max[\{\Delta_m \mid m\neq k \}]$\\
     ~~~$\mu_k(n) \leftarrow \mu_0e^{-2\Delta/f}$\\
     ~~~$\mathbf{g}_k(n) \leftarrow -\mu_k(n)\big[\boldsymbol{\nabla}_k(n) + \sum_{m=1,m\neq k}^K\boldsymbol{\nabla}_m(n-\Delta_m)*c_{mk}(n)\big]$ \\
     ~~~$\mathbf{v}_k(n) \leftarrow \mathbf{w}_k(n)-\mathbf{w}_k(n-1)$ \\
     ~~~$\rho_k(n) \leftarrow \frac{\mathbf{g}^\mathrm{T}_k(n)\mathbf{v}_k(n)}{||\mathbf{g}_k(n)||\;||\mathbf{v}_k(n)||}$ \\
     ~~~$\beta_k(n) \leftarrow \text{min}(\beta_0|\rho_k(n)|^p,\beta_0)$ ~~ $(0<p<1)$\\
     ~~~$\mathbf{w}_k(n+1) \leftarrow \mathbf{w}_k(n) + \mathbf{g}_k(n) + \beta_k(n)\mathbf{v}_k(n)$\\
     /*Output control signal for the secondary source*/\\
     ~~~$y_k(n) \leftarrow \mathbf{w}_k^\mathrm{T}(n)\mathbf{x}(n) $\\
     /*Calculate local gradient and send to other nodes*/\\
     ~~~$\boldsymbol{\nabla}_k(n) \leftarrow [\mathbf{x}(n)*\hat{s}_{kk}(n)]\cdot e_k(n) $  ~~~~~~~~ $\triangleright$ Send to other nodes\\
     \textbf{end while}\\
    \hline
    \end{tabular}
\end{table}

\section{Numerical simulations}
\label{sec:sims}
In this section, numerical simulations are conducted to evaluate the performance of AMAS-MGDFxLMS algorithm. The primary and secondary acoustic paths are measured in a real noise chamber equipped with an ANC window. The system configuration follows the setup in \cite{Ji2025MGDFXLMS}, consisting of six ANC nodes. The lengths of the secondary paths, compensation filters, and control filters are set to $256$, $33$, and $512$, respectively. The sampling frequency is $16$ kHz. The primary noise is broadband noise ranging from $100$ to $1,000$Hz. The step size is chosen as $1\times10^{-7}$ and $p$ is $0.25$ for all the simulations. The proposed AMAS-MGDFxLMS is compared with MGDFxLMS, ASSS-MGDFxLMS and fixed momentum ASSS-MGDFxLMS (FMAS-MGDFxLMS). To quantitatively assess noise reduction performance, the average normalized squared error (ANSE) across all ANC nodes \cite{Ji2025MGDFXLMS}.

\subsection{Noise reduction performance under sudden changes in communication delay}
\begin{figure}[!t]
    \centering
    \includegraphics[width = 0.8\columnwidth]{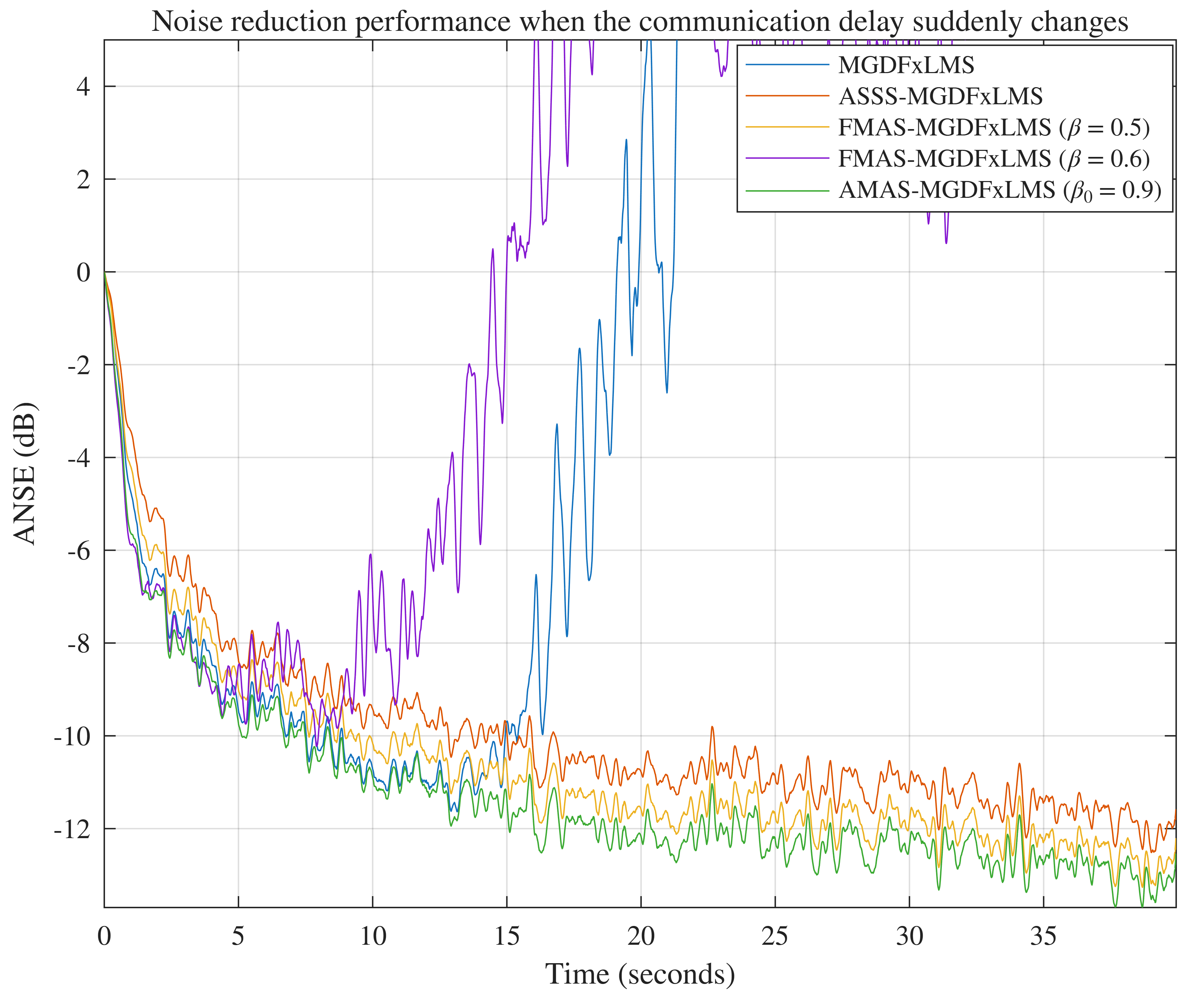}
    \caption{Noise reduction performance using different algorithms when communication delay suddenly changes at the 10s, 20s, and the 30s.}
    \label{fig:case1}
\end{figure}
\begin{figure}[!t]
    \centering
    \includegraphics[width = 0.8\columnwidth]{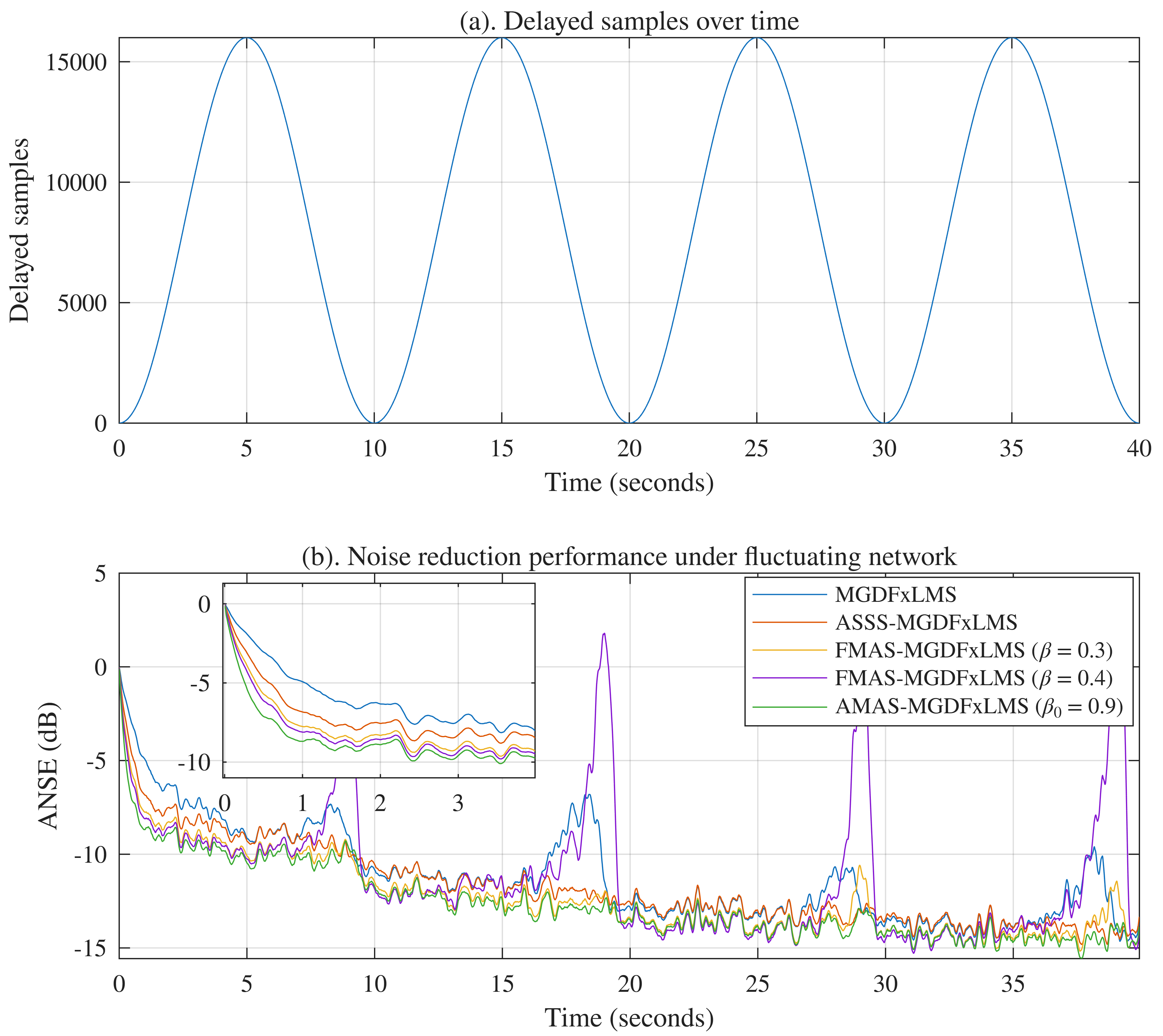}
    \caption{(a). Fluctuating network where communication delay gradually changes over 0-1 seconds; (b). Noise reduction performance using different algorithms under fluctuating network}
    \label{fig:case2}
\end{figure}

In this simulation, the communication delay undergoes variations at the 10s (from the initial 4,000 delayed samples to 8,000 samples), the 20s (from 8,000 samples to 16,000 samples), and the 30s (from 16,000 samples to 8,000 samples). As shown in Fig.~\ref{fig:case1}, when the communication delay changes abruptly, the MGDFxLMS algorithm diverges due to the outdated gradient information. In contrast, the ASSS-MGDFxLMS algorithm preserves convergence stability at the cost of a reduced convergence speed, which motivates the introduction of a momentum term. However, the FMAS-MGDFxLMS algorithm is highly sensitive to the choice of the momentum factor. A large factor leads to over-acceleration and divergence, whereas a small factor fails to provide effective acceleration. To address this issue, the proposed AMAS-MGDFxLMS algorithm adaptively adjusts the acceleration rate according to the directional consistency between the mixed gradients and momentum terms, enabling the DMCANC system to achieve fast noise reduction with maximal acceleration while maintaining stability.

\subsection{Noise reduction performance under fluctuating network}

The network latency is subject to fluctuations in practice. In order to further illustrate the efficacy of the proposed algorithm, we presuppose that the network latency undergoes a gradual transition from $0$ to $1$ seconds, as shown in Fig.~\ref{fig:case2}(a). It can be observed from Fig.~\ref{fig:case2}(b) that the MGDFxLMS algorithm still exhibits instability under fluctuating network conditions. In contrast, the ASSS-MGDFxLMS effectively mitigates this issue and enables stable convergence of the DMCANC system. Although introducing a momentum term can accelerate convergence, an improperly chosen momentum factor may also introduce additional instability. Compared with the FMAS-MGDFxLMS approach, the proposed AMAS-MGDFxLMS dynamically adjusts the momentum factor, achieving a favorable balance between convergence speed and robustness while avoiding the difficulty of manual momentum factor tuning. As a result, the AMAS-MGDFxLMS attains faster and more stable noise reduction performance, demonstrating practical significance.

\section{Conclusion}
\label{sec:concl}
This paper introduces an adaptive momentum mechanism into the ASSS-MGDFxLMS algorithm, resulting in the proposed AMAS-MGDFxLMS algorithm. Building upon the inherent stability of ASSS-MGDFxLMS under varying communication delays and its ability to achieve satisfactory noise reduction, the proposed method further enhances performance by incorporating adaptive momentum. By evaluating the directional consistency between the mixed-gradient updates and the momentum term, the momentum factor is dynamically adjusted to provide maximal acceleration when beneficial while preserving system stability. Therefore, the proposed AMAS-MGDFxLMS achieves faster convergence and improved noise reduction performance under communication-constrained conditions, demonstrating strong practical significance for real-world DMCANC applications.

\section*{Acknowledgment}
This work was supported by the Ministry of Education, Singapore, through Academic Research Fund Tier 2 under Grant MOE-T2EP50122-0018.

\bibliographystyle{IEEEtran}
\bibliography{refs}

\end{document}